\documentclass[twocolumn,showpacs,amsmath,amssymb]{revtex4}
\usepackage{amssymb}
\usepackage[dvips]{graphicx}
\begin{document}

\title{Angle-resolved photoemission spectroscopy of band tails  in  lightly doped cuprates}
\author{A. S. Alexandrov and K. Reynolds}

\affiliation{Department of Physics, Loughborough University,
Loughborough LE11 3TU, United Kingdom\\}

\begin{abstract}
We amend  \emph{ab initio} strongly-correlated band structures by
taking into account the band-tailing phenomenon in  doped
charge-transfer Mott-Hubbard insulators. We show that  the
photoemission from band tails accounts for sharp "quasi-particle"
peaks, rapid loss of their intensities in some directions of the
Brillouin zone ("Fermi-arcs")
 and high-energy "waterfall" anomalies  as a
consequence of matrix-element effects of disorder-localised states
in the charge-transfer gap of doped cuprates.

\end{abstract}


\maketitle

 Since the discovery of high-$T_c$ superconductivity in
cuprates, angle-resolved photoemission spectroscopy (ARPES) has
offered a tremendous advance into the understanding of their
electronic structure  \cite{shen}. However, even though ARPES is
continually strengthening our insights into the band structure and
correlations in cuprates, it has also revealed many poorly
understood phenomena, such as the incoherent "background", the sharp
"quasi-particle" peaks  near some points of the Brillouin zone (BZ),
which form "arcs" of "Fermi surface" (FS) (\cite{yoshida} and
references therein), widely studied low-energy dip-hump and
 kink features (for review see \cite{shen}) and the more
recently discovered steep downturn of the dispersion toward higher
energies (the so-called "waterfall")
\cite{ronning,graf,meevasana,xie,pan,chang,kordyuk}. These anomalies
have received quite different interpretations, involving, for
example, uncorrelated \cite{aledyn} and strongly-correlated
\cite{fehske,nagaosa,ronning,wolfgang,gun} lattice polarons,
Migdal-Eliashberg-like approaches \cite{hag,maks}, spinons and
holons \cite{graf}, spin polarons \cite{xie},  spin fluctuations
\cite{bor,mac} and band-structure matrix element effects
\cite{bansil,kordyuk}.

Many  ARPES interpretations suggest a large FS (as an
 exception see e.g. \cite{aledyn}) with  nodal  gapless
quasiparticles, which are gapped or strongly damped in the antinodal
directions ($(0,0) \rightarrow (\pi,0)$) of the two-dimensional (2D) BZ.
Importantly, extensive simulations  of ARPES using the
first-principles (LDA) band theory with the matrix elements
properly taken into account \cite{bansil} reproduces well the
topological features of momentum distribution curves (MDC), pointing
to the large FS in optimally doped cuprates. However, LDA predicts that the undoped
parent cuprates are metallic with
roughly the same large FS, while they are actually charge-transfer
Mott-Hubbard insulators with the optical gap at 2 eV. This fact led
to  several powerful extensions of LDA, in particular to LDA+U, which
combines LDA eigenfunctions with strong Coulomb correlations
introduced as a model parameter (Hubbard U) \cite{zaanen}, and the
LDA+generalized tight-binding (GTB)
 method combining the exact diagonalization of the
intracell part of the Hamiltonian with relevant LDA eigenfunctions
and Coulomb correlations  and the perturbation treatment of the
intercell hoppings and interactions \cite{kor}. LDA-GTB Hamiltonian
is reduced to the simpler effective $t-J$ or $t-J^*$ model ($t-J$
model plus three-center correlated hoppings \cite{kor}) in the
low-energy domain.

LDA+GTB band structure of undoped cuprates with \emph{ab initio}
sets of tight-binding parameters \cite{kor} describes remarkably
well the optical gap, $E_{ct} \approx 2$ eV both in
antiferromagnetic and paramagnetic states of the undoped
La$_2$CuO$_4$. The valence band consists of a set of very narrow
($\lesssim$ 1 eV) subbands where the highest one is dominated by the
oxygen $p$ states with the maximum at ${\bf k}\equiv {\bf g}=
(\pi/2a,\pi/2a)$  (see Fig.1), while the bottom of the empty
conduction band formed by $d_{x^2-y^2}$ states of copper is found at
$(\pi/a,0)$. These locations of valence-band maximum and
conduction-band minimum perfectly agree  with ARPES intensity locus
in hole doped La$_{2-x}$Sr$_x$CuO$_4$  and electron-doped
Nd$_{2-x}$Ce$_{x}$CuO$_4$, respectively \cite{yoshida2}. Importantly, the
LDA+GTB approach predicts the charge-transfer gap at any doping with
the chemical potential pinned near the top of the valence band (in
hole doped cuprates) and near the bottom of the conduction band (in
electron-doped cuprates)  due to
 spin-polaron $in-gap$ states.

  ARPES
 of undoped  cuprates \cite{yoshida,yoshida2,shen,ronning,kshenscience,kshen2007} proved to be critical
 in the assessment of different theoretical approaches. It revealed an
 apparent contradiction with the $t-J$ model. There is no sharp peak predicted by the model in
undoped cuprates,  but  a slightly dispersive broad incoherent
background, Fig.2 (inset). Small lattice polarons due to a  strong
electron-phonon interaction (EPI) have been advocated as a plausible
explanation of the discrepancy \cite{nagaosa}. When EPI is strong,
the  spectral weight, $Z$, of the coherent small-polaron peak is
very small, $Z\ll 1$  and, hence the peak can not be seen in
experiment since all weight of the sharp resonance in the $t-J$
model is transformed at strong EPI into the broad continuum.

\begin{figure}
\begin{center}
\includegraphics[angle=-90,width=0.50\textwidth]{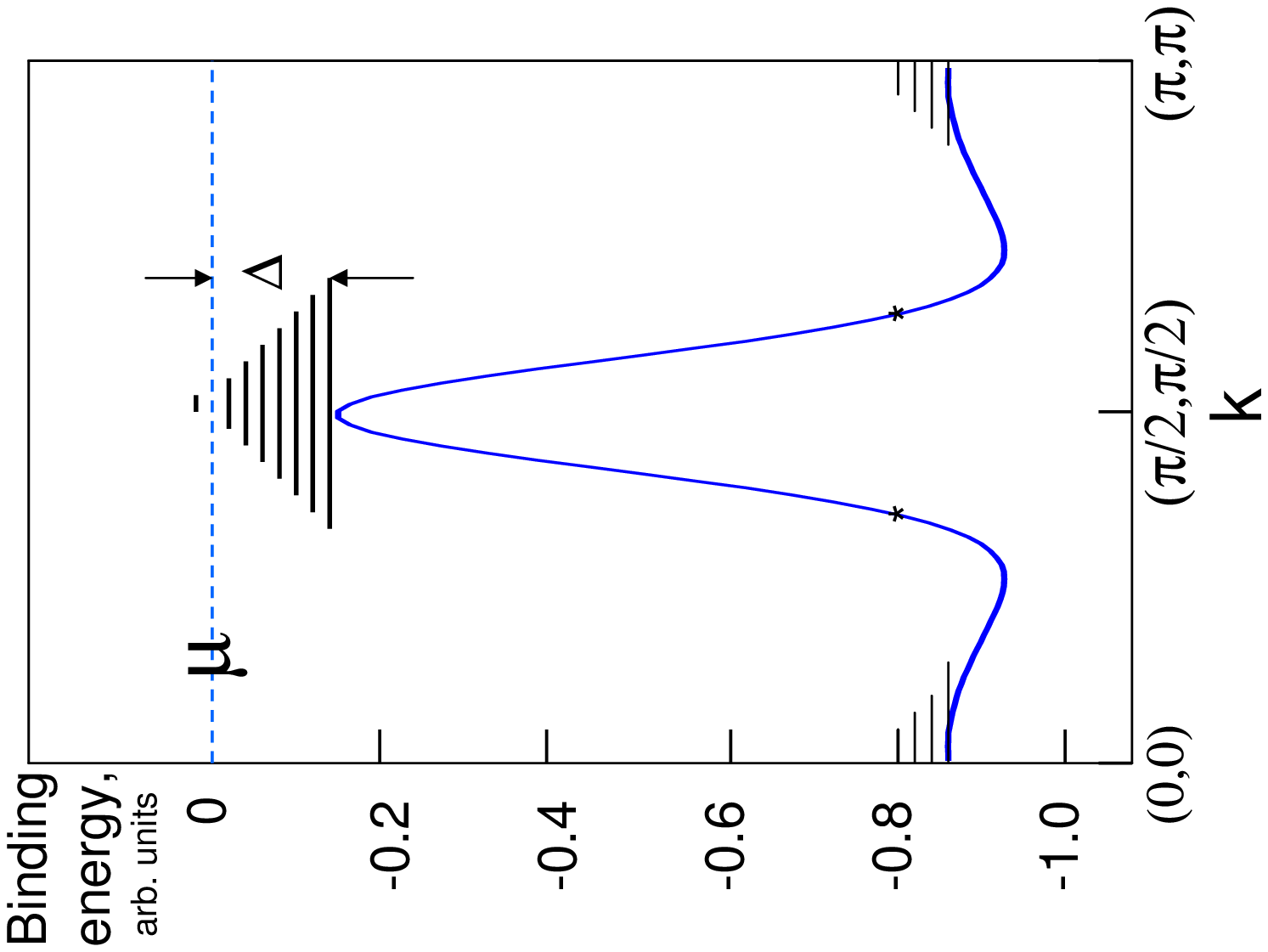}
\vskip -0.5mm \caption{LDA+GTB valence band dispersion \cite{kor}
amended with  band tails (ladder lines) near $\Gamma$,
$(\pi/2,\pi/2)$ and $(\pi,\pi)$ maxima ( here k is measured in
$1/a$)}
\end{center}
\end{figure}

Unfortunately the  energy distribution curves (EDC)in
La$_{2}$CuO$_4$ , Fig.2a, has only little if any resemblance to the
small-polaron spectral function, which is roughly gaussian-like.
Only by subtracting a "background" given by the spectrum near
($\pi/a,\pi/a $), Fig.2a, one can account for the remaining EDC with
the polaronic spectral function \cite{gun}. This background problem
obscures any reliable interpretation of the broad ARPES intensities,
especially in underdoped cuprates, where the charge-transfer gap at
2 eV makes inelastic scattering events implausible as an explanation
of the
 background.  Sharp peaks at $(\pi/2a,\pi/2a)$ near the
Fermi level, Fig.2b, in doped cuprates also  remains a puzzle. Small
heavy polarons cannot screen EPI  in lightly doped cuprates. Hence,
if $Z$ is small in the parent cuprate, it should also remain small
at finite doping, so that the emergency of the peaks cannot be
explained by a substantial increase of $Z$ with doping.

Here we show that amending  the LDA+GTB band structure of doped
cuprates by inevitable  impurity band-tails, the ARPES puzzles as
mentioned above are explained.

Doping of cuprates inserts a large number of impurities into the
parent lattice. Each impurity ion locally introduces a distinct
level, $E_i$, in the charge-transfer gap. The fact that the
impurities are randomly distributed in space causes the density of
states (DOS) to tail, like in heavily doped semiconductors
\cite{mieghem}. When there are many impurities within the range
$\xi_i$  of a localised wave function $\psi_i({\bf r})$, the random
potential produces low-energy states near maxima of the valence band
at hole doping, Fig.1, or near minima of the conduction band at
electron doping. As a result, ARPES intensity, $I({\bf k},
E)=I_{b}({\bf k},E)+I_{im}({\bf k},E)$ comprises the band-tail
intensity, $I_{im}({\bf k},E)$, due to localised states within the
charge-transfer gap, and the valence band contribution, $I_{b}({\bf
k},E)$, of
 itinerant Bloch-like states. According to LDA band structures
\cite{bansil} the itinerant states are anisotropic-3D (specifically
in La$_2$CuO$_4$) dispersing with c-axis  $k_z$ over a few hundred
meV.  We suggest that  this dispersion shapes the background making
it so different from the incoherent background caused by EPI and/or
spin fluctuations since $k_z$ is not conserved in ARPES experiments.
On the other hand    the incoherent background can be well described
by a simple polaronic Gaussian in presumably more anisotropic
 insulating Ca$_2$CuO$_2$Cl$_2$\cite{kshen2007}.

Here we focus on the band-tailing contribution described by the
Fermi-Dirac golden rule as
\begin{equation}
I_{im}({\bf k},E)={2\pi e^2\over{m_{e}^2}}n(E) \sum_{i}
\left|\left\langle\psi_{f} \left|{\bf A}_0\cdot {\bf
\nabla}\right|\psi_{i}\right\rangle\right|^2\delta(E+\Delta-E_i).
\end{equation}
We define all energies relative to the chemical potential, $\mu$,
which is situated within the impurity band as shown in Fig.1. Only
the impurity states with the binding energy $E_i$  below $\mu=0$
contribute at zero temperature. Here ${\bf A}_0$ is the amplitude of
X-ray vector potential, and $\hbar=k_B=1$.

We take  the impurity  wavefunction  as \cite{kohn}, $\psi_i({\bf
r})=F_i({\bf r})\psi_{\bf g}({\bf r})$, and the final state to be
the normailsed plane wave, $\psi_f({\bf r})=(Nv)^{-1/2}\exp(i{\bf k
\cdot r})$. Here $\psi_{\bf g}({\bf r})$ is the itinerant state at
the top of the valence band, $F_i({\bf r})$ is a slowly varying
envelope function, and $N$ is the number of unit cells of volume $v$
in the crystal. In the framework of GTB \cite{kor} one can expand
$\psi_{\bf g}({\bf r})$ using the Wannier orbitals, $\psi_{\bf
g}({\bf r})=N^{-1/2} \sum_{\bf m} w({\bf r-m}) \exp(i{\bf g \cdot
m})$, and calculate the dipole matrix element in Eq.(1) as
\begin{equation}
I_{im}({\bf k},E)=I n(E) \sum_{i} |f_i({\bf
k-g})|^2\delta(E+\Delta-E_i),
\end{equation}
where $I=2\pi(ed/m_e)^2({\bf A}_0 \cdot {\bf k})^2/v$ is
proportional to the valence band matrix element squared, which is
roughly a constant in a wide range of $\bf k$ near $\bf g$, $d=\int
d{\bf r}w({\bf r}) \exp(i{\bf g \cdot r})$, and $f_i({\bf
q})=(Nv)^{-1}\int d{\bf r} \exp(i{\bf q \cdot r}) F_i({\bf r})$ is
the Fourier transform of the impurity envelope function.

Since the size of the envelope is large compared with the lattice
constant, its Fourier transform strongly depends on ${\bf q}$, which
explains the experimental  EDC and MDC as we show in the rest of the
paper. We choose the impurity state to be hydrogen-like, $F_i({\bf
r})=(Nv/\pi \xi_i^3)^{1/2} \exp(-r/\xi_i)$ as the hydrogen model
accurately predicts many properties of shallow levels in heavily
doped semiconductors, so that
 $f_i({\bf q})=8\pi (\xi_i^3/\pi Nv)^{1/2}
 (1+q^2\xi_i^2)^{-2}$
for  3D impurity states, and $f_i({\bf q})\propto
 (1+q^2\xi_i^2)^{-3/2}$ for 2D states like localised surface states.
 It is important to recognise here that $\xi_i$ is related to
the impurity binding energy as $\xi_i^{-2}=mE_i$, where $m$ is
roughly the hole effective mass. As a result we  get $I_{im}({\bf
k},E)=xIn(E)M({\bf k-g}, E)$ with
\begin{equation}
M({\bf k-g}, E)={64 \pi\over{vm^{3/2}}}
{(E+\Delta)^{5/2}\over{[E+\Delta+({\bf k-g})^2/m]^{4}}}
\rho_{im}(E+\Delta).
\end{equation}
Here $\rho_{im}(E)=N_i^{-1}\sum_{i} \delta(E-E_i)$ is the band-tail
density of states (DOS) normalised to unity, and  $x=N_i/N$ is the
impurity concentration per cell proportional to  doping. In the 2D
case the result is similar, $M_{2D}({\bf k}, E)\propto
(E+\Delta)^{2}[E+\Delta+({\bf
k_{\parallel}-g})^2/m]^{-3}\rho_{im}(E+\Delta)$.

\begin{figure}
\begin{center}
\includegraphics[angle=-90,width=0.55\textwidth]{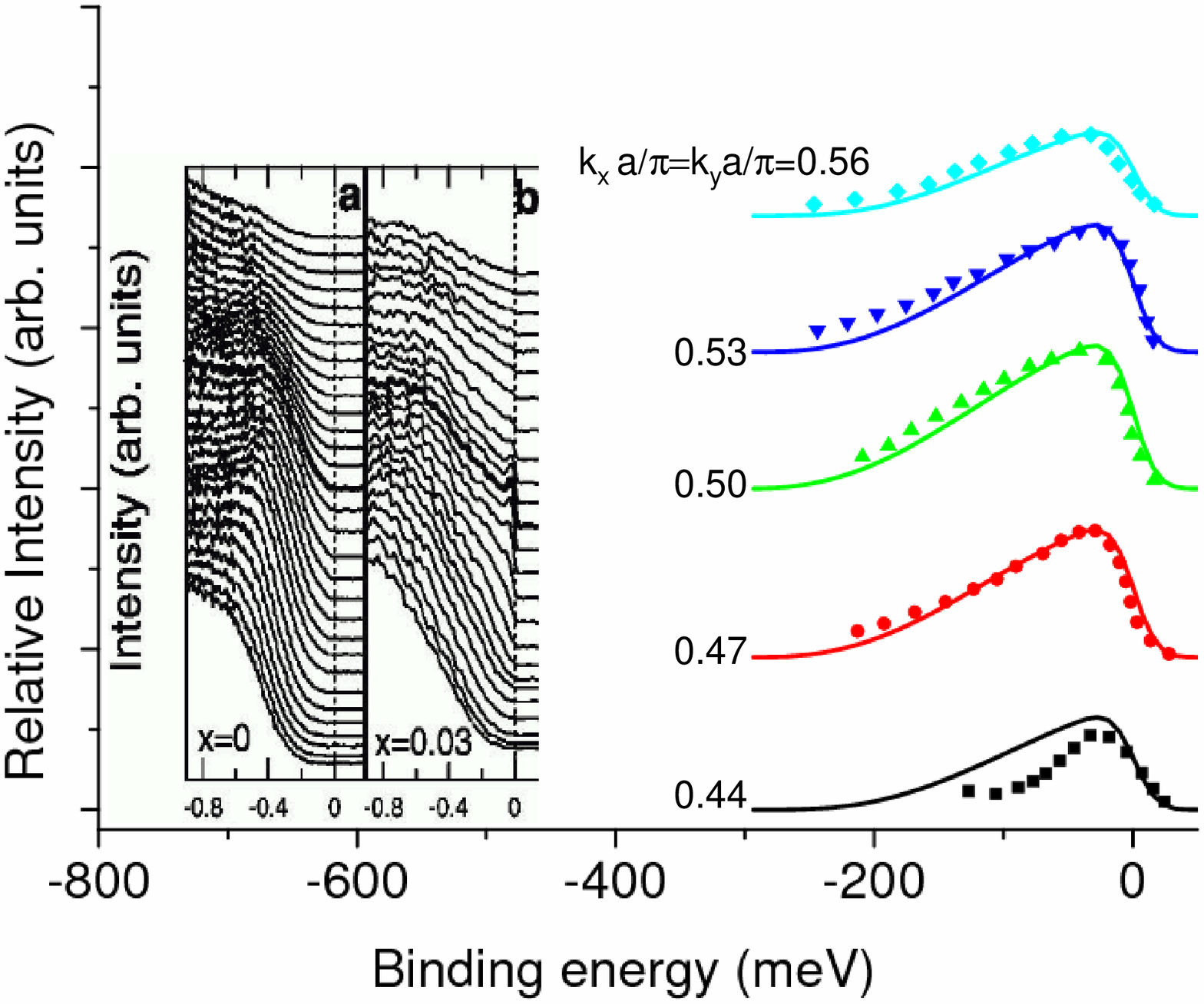}
\vskip -0.5mm \caption{Band-tail EDC, Eq.(4), (solid lines) with
pseudogap $\Delta=300$ meV and  band-tail width $\gamma=300$ meV
compared with  relative EDC (symbols) near $(\pi/2a,\pi/2a)$.
Relative intensities are obtained by subtracting ARPES intensities
of the parent compound, La$_2$Cu0$_4$ (a), shifted by $\delta \mu$,
from EDC of slightly doped La$_{1.97}$Sr$_{0.03}$CuO$_4$ (b) as
measured by Yoshida \emph{et al.} \cite{yoshida}. Both intensities
have been normalised by their values at $E=-800$ meV and the
chemical potential shift between two samples has been taken as
$\delta \mu=70$ meV. }
\end{center}
\end{figure}

We notice that due to a very sharp dependence on $q$ of the matrix
element in Eq.(2) any uncertainty of $k_z$ does not smear out the
strong dependence  of $I_{im}({\bf k}, E)$ on the in-plane momentum
component, ${\bf k}_{\parallel}$. Averaging over $k_z$ simply
replaces $M({\bf k-g}, E)$ in Eq.(3) by
\begin{equation}
\tilde{M}({\bf k}_{\parallel}-{\bf g}, E)\approx{32 c \over{vm}}
{(E+\Delta)^{5/2}\over{[E+\Delta+({\bf
k_{\parallel}-g})^2/m]^{7/2}}} \rho_{im}(E+\Delta),
\end{equation}
where $c$ is the c-axis lattice constant. Also $M$ and $\tilde{M}$
can be very large for shallow impurity states, $M, \tilde{M}\gg
1/x$. Hence even the
 strong polaronic reduction of their weight, $Z\ll 1$,
does  not make band-tails invisible in ARPES at finite doping, in
contrast to a complete reduction of the coherent band peak.

\begin{figure}
\begin{center}
\includegraphics[angle=0,width=0.20\textwidth]{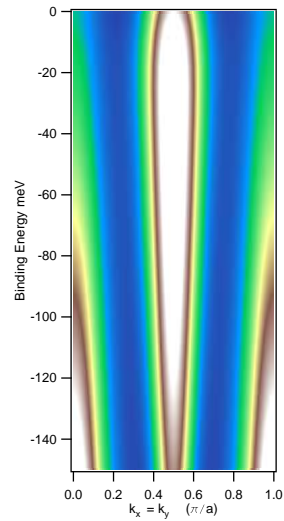}
\vskip -0.5mm \caption{Waterfall effect in the band-tail ARPES
intensity (white colour corresponds to the  highest intensity).}
\end{center}
\end{figure}

 Since the chemical potential shifts
towards the band edge with doping, $\Delta$ in Eqs.(3,4) becomes
smaller. Hence, the band-tail peak, $I_{im}({\bf k},E)$, which is
proportional to $x$, not only increases but also becomes sharper
with doping as observed \cite{yoshida}. To provide more insight into
the shape and momentum dependence of experimental EDC we approximate
the band-tail DOS by the simple form,
$\rho_{im}(E)=[n/\Gamma(p/n+1/n)](E/\gamma)^{p}\exp(-E^n/\gamma^n)$,
where $\Gamma(x)$ is the gamma-function. Exponents $n,p$ depend on
the dimensionality and the correlation length of the disorder
potential: $n=2$ both in 2D and 3D, $p=2$ in 2D and $p=7/2$ in 3D
for the long range random potential correlations. In the short-range
Gaussian-white-noise limit one obtains $n=1,1/2$ in 2D and 3D,
respectively, and $p=3/2$ in both dimensions \cite{hal}. We can
separate impurity and band contributions by subtracting normalised
ARPES intensity  of the parent cuprate from the intensity of the
doped one. Then, the band-tail ARPES, Eq.(4), fits
 well with the experimental relative intensities at all momenta
around ${\bf g}$ with $m=m_e$, $n=2$, and $p=7/2$, Fig.2. It
describes the substantial loss of intensity with changing the
momentum by only a few percent relative to ${\bf g}$, as well as the
shape of the relative EDC.

\begin{figure}
\begin{center}
\includegraphics[angle=-90,width=0.55\textwidth]{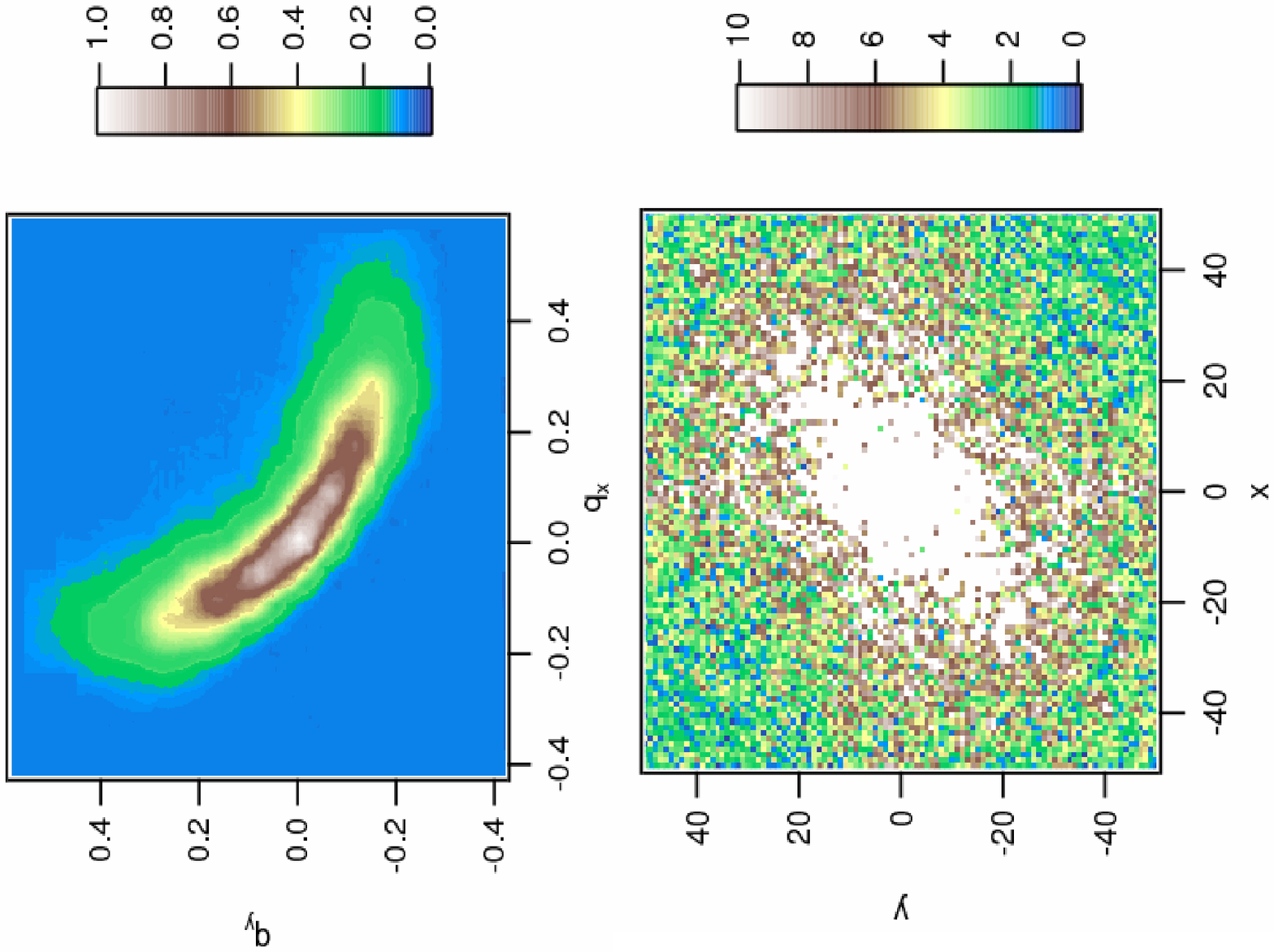}
\vskip -0.5mm \caption{Real space Fourier transform (lower panel) of
the square root of ARPES intensities (arb. units) at the Fermi level
in Ca$_{2-x}$Na$_{x}$CuO$_2$Cl$_2$ (upper panel, measured by Shen
\emph{et al.} \cite{kshenscience} for $x=0.12$) reveals the
real-space size (in units of $a$) of localised in-gap states.}
\end{center}
\end{figure}

We argue that  band-tailing can  also contribute to the waterfall
effect. There are impurity tails near local maxima of the LDA+GTB
valence band at $\Gamma$ point $(0,0)$ and at ${\bf
g}_1=(\pi/a,\pi/a)$, as shown in Fig.1. Different from  in-gap
impurity states at ${\bf g}=(\pi/2a,\pi/2a)$ these localised states
are hybridised with the valence band states of the same energy
(shown by stars in Fig.1). However, the hybridisation could be
insignificant, if the corresponding matrix elements of the random
potential are small due to a large momentum separation between those
states of the order of $\pi/2a$. Hence, the impurity peaks  reappear
and disperse towards $(0,0)$ and ${\bf g}_1$ at high binding
energies, as observed in a number of doped cuprates
\cite{graf,meevasana,xie,pan,chang,kordyuk}. We illustrate the
waterfall in  Fig.3  by adding all three tail contributions,
$I_{im}({\bf k},E)\propto n(E)[\tilde{M}({\bf k}_{\parallel},
E+E_2)+\tilde{M}({\bf k}_{\parallel}-{\bf g}, E)+\tilde{M}({\bf
k}_{\parallel}-{\bf g_1}, E+E_2)]$  where $E_2$ is roughly the
valence band-width (we chose $E_2=500$ meV). We notice that the
Fermi-Dirac distribution, $n(E)$, is replaced by its convolution
with the Gaussian energy resolution function, $n(E)\rightarrow
[1-erf(E/\delta)]/2$ in plotting Figs.2,3 since the energy
resolution $\delta=20$ meV is much larger than $T\approx 2$ meV.
Also the photoemission intensity  comprises both  band-tail and
valence band contributions, so that the resulting dispersion could
be different from the anomalous band-tail dispersion of relative
intensities, Fig. 2.

 Our theory proposes that the ARPES  intensity
near $(\pi/2a,\pi/2a)$ is proportional to the square of the Fourier
component, $f_i ({\bf q})$, of the impurity wave-function envelope,
Eq.(2). Therefore, we can find the real-space image of the function,
$F_i({\bf r})$, by taking the Fourier transform of the square root
of the experimental intensities, Fig.4 (upper panel). Here we show
the intensities near the Fermi level measured in
Ca$_{2-x}$Na${_x}$Cu O$_2$Cl$_2$ \cite{kshenscience}, which are very
similar, if not identical to those in La$_{2-x}$Sr$_{x}$CuO$_4$
(compare Fig.1 \cite{kshenscience} and  Fig.2 in \cite{yoshida}).
The real-space image (lower panel, Fig.4) reveals some band-mass
anisotropy and the size of the localised state  of about 20 lattice
constants justifies the "envelope" approximation \cite{kohn} used
for the impurity wavefunction.

In summary, we have proposed an explanation  for sharp
"quasi-particle" peaks, "Fermi-arcs",
 and the high-energy waterfall in  cuprates as a
consequence of matrix-element effects  of disorder-localised
band-tails in the charge-transfer gap of doped Mott-Hubbard
insulators. Importantly if holes are bound into  bipolarons, the
chemical potential remains   within the single-particle band-tail at
the bipolaron mobility edge even up to optimum doping, in agreement
with $S-N-S$ tunnelling experiments \cite{boz} and insulating-like
low-temperature resistivity of underdoped cuprates. In this case
$\Delta$ in Fig.1 is half of the bipolaron binding energy
\cite{aledyn}, which is also the normal state pseudogap
\cite{alenar}. Recent scanning tunnelling microscopy at the atomic
scale found intense nanoscale disorder in high-Tc superconductor
Bi$_2$Sr$_2$CaCu$_2$O$_{8+\delta}$ \cite{davis} telling us  that
band-tailing indeed plays the important role in shaping
single-particle spectral functions of doped Mott insulators.

We are grateful  to ZX Shen and Teppei Yoshida for providing us with
their raw ARPES data \cite{yoshida} and enlightening comments. We
greatly appreciate valuable discussions with Arun Bansil, Sergey
Borisenko, Ivan Bozovic, Jim Hague, Jan Jung, Alexander  Kordyuk,
Maxim Korshunov, Kyle Shen, and Jan Zaanen. This work was supported
by EPSRC (UK) (grant number EP/C518365/1).


\begin{thebibliography}{90}

\bibitem{shen} A. Damascelli, Z. Hussain and Zhi-Xun Shen, Rev. Mod. Phys. {\bf 75} 473
(2003);  X. J. Zhou \emph{et al.}, cond-mat/0604284.
\bibitem{yoshida} T. Yoshida \emph{et al.},
Phys. Rev. Lett. {\bf 91}, 027001 (2003).
\bibitem{ronning} F. Ronning \emph{et al.}, Phys. Rev. B {\bf 71}, 094518 (2005).
\bibitem{graf} J. Graf \emph{et al.}, Phys. Rev. Lett. {\bf 98}, 067004 (2007).
\bibitem{meevasana} W. Meevasana \emph{et al.}, cond-mat/0612541.
\bibitem{xie} B. P. Xie \emph{et al.},
cond-mat/0607450
\bibitem{pan} Z.-H. Pan \emph{et al.}, cond-mat/0610442.
\bibitem{chang} J. Chang \emph{et al.}, cond-mat/0610880.
\bibitem{kordyuk} A. A. Kordyuk \emph{et al.}, cond-mat/0702374.
\bibitem{mac}  A. Macridin \emph{et al.}, cond-mat/0701429.
\bibitem{aledyn} A. S. Alexandrov and C. J. Dent, Phys. Rev. B {\bf 60},
15414 (1999); A. S. Alexandrov and C. Sricheewin, Europhys. Lett.
{\bf 58}, 576 (2002).

\bibitem{fehske} G. Wellein, H. Roder, and H. Fehske, Phys. Rev. B {\bf 53}, 9666 (1996)
\bibitem{nagaosa} A. S. Mishchenko and N. Nagaosa, Phys. Rev. Lett. {\bf 93}, 036402
(2004).
\bibitem{wolfgang} M. Hohenadler\emph{et al.} , Phys. Rev. B {\bf 71}, 245111 (2005).
\bibitem{gun} O. Rosch \emph{et al.},  Phys. Rev. Lett.
{\bf 95}, 227002 (2005).
\bibitem{hag} J. P. Hague,  J. Phys.: Condense Matter {\bf 15}, 2535 (2003).
\bibitem{maks} E. G. Maksimov, O. V. Dolgov, and M. L. Kulic, Phys. Rev. B {\bf
72}, 212505 (2005).
\bibitem{bor} S. V. Borisenko \emph{et al.}, Phys. Rev. Lett. {\bf 96}, 117004
(2006).
\bibitem{bansil} M. Lindroos, S. Sahrakorpi, and A. Bansil, Phys. Rev. B {\bf 65}, 054514
(2002).


\bibitem{zaanen} V. I. Anisimov, J. Zaanen, and O. K. Andersen, Phys. Rev. B {\bf 44}, 943
(1991).

\bibitem{kor} S. G. Ovchinnikov \emph{et al.}, J. Phys.: Condens. Matter {\bf 16}, L93 (2004);
M. M. Korshunov \emph{et al.}, Phys. Rev. B {\bf 72}, 165104 (2005).

\bibitem{yoshida2} N. P. Armitage \emph{et al.}, Phys. Rev. Lett. {\bf 88}, 257001 (2002);
K. M. Shen \emph{et al.},
Phys. Rev. B {\bf 69}, 054503 (2004).

\bibitem{kshenscience} K. M. Shen \emph{et al.},
Science {\bf 307}, 901 (2005).


\bibitem{kshen2007} K. M. Shen \emph{et al.},
Phys. Rev. B {\bf 75}, 054503 (2007).

\bibitem{mieghem} P. V. Mieghem, Rev. Mod. Phys. {\bf 64}, 755
(1992).

\bibitem{kohn}
W. Kohn and J. M. Luttinger, Phys. Rev. ${\bf 97}$, 869-883 (1955).

\bibitem{hal} B. I. Halperin and M. Lax, Phys. Rev. {\bf 148}, 722
(1966); R. Eymard and G. Duraffourg, J. Phys. D: Appl. Phys. {\bf
6}, 66 (1973); D. N. Quang and N. H. Tung, Phys. Stat. Sol. B {\bf
209}, 375 (1998).


\bibitem{boz}
I. Bozovic \emph{et al.},  Nature (London) {\bf 422}, 873 (2003).

\bibitem{alenar}  A. S. Alexandrov, in \emph{Studies in High Temperature
Superconductors,} ed. A.V.  Narlikar (Nova Science Pub., NY 2006),
\textbf{50}, pp. 1-69.
\bibitem{davis} J.  Lee \emph{et al.},  Nature (London), {\bf 442} 546 (2006).





\end{thebibliography}
\end{document}